# Near-tropical subsurface ice on Mars

Mathieu Vincendon[1], John Mustard[1], François Forget[2], Mikhail Kreslavsky[3], Aymeric Spiga[4], Scott Murchie[5] & Jean-Pierre Bibring[6]

[1]Department of Geological Sciences, Brown University, Providence, RI, USA (mathieu.vincendon@ias.u-psud.fr). [2]Laboratoire de Météorologie Dynamique, Université Paris 6, Paris, France. [3]Earth and Planetary Sciences, University of California - Santa Cruz, CA, USA. [4]Department of Physics & Astronomy, Open University, Milton Keynes, UK. [5]Applied Physics Laboratory, Johns Hopkins University, Laurel, MD, USA. [6]Institut d'Astrophysique Spatiale, Université Paris Sud, Orsay, France.

**Abstract:** Near-surface perennial water ice on Mars has been previously inferred down to latitudes of about 45° and could result from either water vapor diffusion through the regolith under current conditions or previous ice ages precipitations. In this paper we show that at latitudes as low as 25° in the southern hemisphere buried water ice in the shallow (< 1 m) subsurface is required to explain the observed surface distribution of seasonal $CO_2$ frost on pole facing slopes. This result shows that possible remnants of the last ice age, as well as water that will be needed for the future exploration of Mars, are accessible significantly closer to the equator than previously thought, where mild conditions for both robotic and human exploration lie.

While perennial water ice is routinely observed both at the surface and in the subsurface at high latitudes [*Kieffer et al.*, 1976; *Boyton*, 2002; *Mitrofanov*, 2002; *Bibring et al.*, 2004; *Mellon et al.*, 2004; *Bandfield and Feldman*, 2008; *Smith et al.*, 2009], only subsurface water ice can survive throughout the entire year at mid-latitudes. Evidence for shallow (< 1 m) subsurface water ice has been obtained from observations down to latitudes of about 45° in both hemispheres [*Mellon et al.*, 2004; *Byrne et al.*, 2009], and deeper buried glaciers have been locally inferred down to 40° latitude [*Holt et al.*, 2008; *Plaut et al.*, 2009]. It is not clear yet to what extent this subsurface ice has formed under current conditions via water vapor diffusion through the regolith and/or is the remnant of previous ice ages precipitations that occurred at higher obliquities [*Mellon and Jalosky*, 1995; *Head et al.*, 2003; Schorghofer 2007; *Hudson et al.*, 2009]. At more equatorward latitudes, morphological observations consistent with the past presence of shallow subsurface water ice have been reported [*Squyres and Carr*, 1986; *Mustard et al.*, 2001; *Head et al.*, 2003]. It has been suggested that this water ice may be locally preserved [*Christensen*, 2003], and modeling predictions indicate that subsurface water ice could be stable today on pole facing slopes at those latitudes [*Aharonson and Schorghofer*, 2006]. However, there is no observational evidence.

The OMEGA (*Observatoire pour la Minéralogie, l'Eau, les Glaces et l'Activité* onboard Mars Express) and CRISM (*Compact Reconnaissance Imaging Spectrometer for Mars* onboard Mars Reconnaissance Orbiter) near-infrared imaging spectrometers have been observing the surface of Mars since 2004 and 2006 respectively. These instruments measure solar radiation scattered by the surface, mainly in the first upper hundreds of microns. They provide spectral images with a spatial resolution ranging from 20 meters to 5 kilometers and a spectral sampling between 7 nm and 40 nm. While these data have been widely used to assess the composition of both minerals and condensates on the surface of Mars, they do not provide direct evidence of the properties of the subsurface. Nevertheless, surface conditions are partly driven by the thermal properties of the subsurface. In fact, the current amount of $CO_2$ that condenses as seasonal caps is controlled by the presence of subsurface water ice: high-latitudes water ice in the shallow subsurface is required for global climate models to match the surface pressure data [*Haberle et al.*, 2008]. Condensed $CO_2$ deposits at the surface can be detected in OMEGA and CRISM data using the 1.43 µm absorption band [*Langevin et al.*, 2007]. The band depth detection limit at this wavelength is 2% under ideal conditions and 10% under less ideal conditions. This detection limits correspond to a photon path length within the ice on the order of 10-200 µm [*Schmitt et al.*, 2004]. Most $CO_2$ ice deposits are therefore detectable using OMEGA and CRISM according to their expected thickness and grain size [*Forget et al.*, 1995].

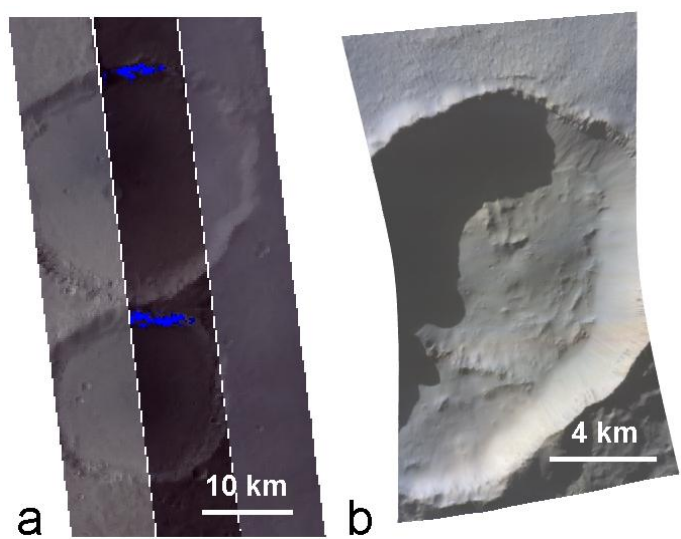

*Figure 1: Near-IR observations of crater slopes in southern winter by the CRISM instrument (North is up). (a) Examples of pole-facing slopes with $CO_2$ ice deposits. A low resolution (230 m/pix) mosaic of two craters at 40°S, 157°E shows $CO_2$ ice (blue) in mid-winter ($L_S$ = 147°, middle observations) but not in early spring ($L_S$ = 200°, left) and late winter ($L_S$ = 165°, right). The spatial and seasonal distributions of these $CO_2$ deposits are shown in Figure 2 and 3. (b) Steep pole facing slope (33°) without ice observed at $L_S$ 88° (winter solstice) in a high resolution observation (34 m/pix) centered at 32.6°S, 131.2°E. The lack of $CO_2$ frost at this latitude and season can only be explained by the presence of subsurface water ice that release during winter the heat accumulated in summer.*

$CO_2$ ice condenses as a seasonal cap between the pole and about 45° latitude in the southern hemisphere [*Langevin et al.*, 2007]. However, patches of $CO_2$ ice accumulated thanks to local thermal conditions have been reported at more equatorward latitudes [*Schorghofer and Edgett*, 2006; *Langevin et al.*, 2007] (Figure 1). To better understand these conditions we have monitored the presence of surface $CO_2$ ice in the southern hemisphere using the OMEGA and CRISM datasets. The spatial and temporal distributions of the detected deposits are shown on Figure 2 and 3 respectively. $CO_2$ ice is detected in late southern fall and winter. The latitudinal stability limit of $CO_2$ ice varies with longitude, with a minimum value of 34°S. $CO_2$ ice is less stable at high altitudes (e.g. in Thaumasia) as expected from the decrease of the frost point temperature with elevation. Our observational results generally agree with previous detection of transient bright deposits in visible images which where inferred to be $CO_2$ ice deposits [*Schorghofer and Edgett*, 2006]. A notable difference is the detection of $CO_2$ ice 10° of solar longitude earlier in the season. Another difference between our study and these results occurs in the Thaumasia regions (~ 270°E) where bright surface ice deposits are composed of water ice and not of $CO_2$ ice according to our near-IR spectroscopy measurements.

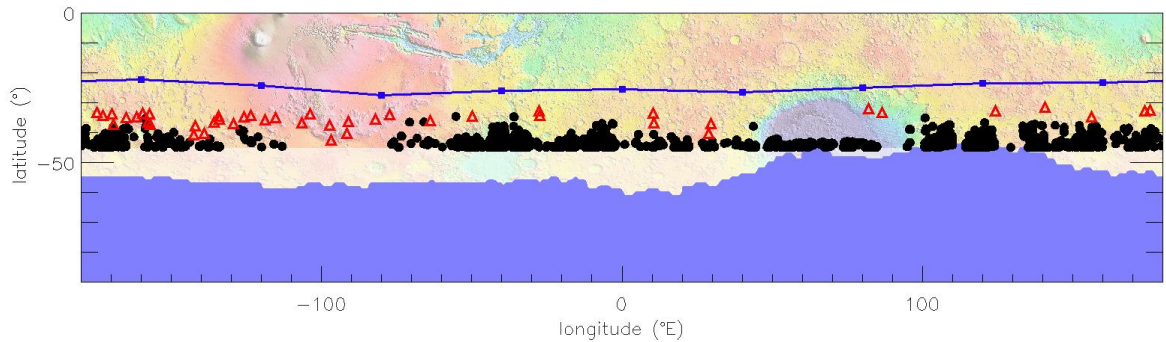

*Figure 2: $CO_2$ ice deposits (black points) observed with CRISM and OMEGA at latitudes higher than 45°S are shown on a MOLA altimetry map [Smith et al., 1999]. $CO_2$ ice is detected up to 34°S. Most $CO_2$ deposits are observed on pole facing slopes with angles in the 20-30° range according to topography measurements [Smith et al., 1999; Kreslavsky and Head, 2000]. Deposits observed at the lowest latitudes are in the upper part of that range. Steep (> 25°) pole facing crater rim without ice (Figure 1b) observed between $L_S = 100°$ and $L_S = 140°$ (when ice is stable between 36°S and 34°S) at latitudes southern than 32°S are also indicated (red triangles). They show that no major observational bias occur in our derived spatial stability limits (except in Hellas due to clouds). The previously well-accepted limit of near-surface ground ice as constrained by low-resolution observations [Mellon et al., 2004] is indicated as a blue-filled area (southern of a limit that range from 60°S to 45°S). Our derived local limit below steep pole facing slopes is indicated as a blue line (around 25°S): poleward of 25°S, water ice is needed in the first meter of the subsurface to explain the low stability of $CO_2$ ice as a function of latitude and season (see Figure 3).*

We have compared this observed stability pattern with the predictions of a one-dimensional energy balance code derived from the Global Climate Model developed at the LMD [*Forget*, 1999; *Spiga and Forget*, 2008]. This model has been extensively validated through comparisons with available spacecraft observations. It calculates the energy balance between the incoming energy fluxes on pole facing slopes (direct sunlight, light scattered and IR radiation emitted and scattered by aerosols and $CO_2$ gas, solar and thermal radiations reflected or emitted by surrounding flat surfaces, heat conducted from the subsurface, and latent heat release when $CO_2$ condenses) and outgoing energy fluxes (thermal IR radiation emitted by the surface, heat conducted into the subsurface, latent heat used for $CO_2$ sublimation). The model also accounts for the sensible heat exchange between the atmosphere and the surface (a minor term due to the low atmospheric pressure). $CO_2$ condensation starts when the surface temperature reaches the $CO_2$ frost point. The presence of $CO_2$ ice then modifies the surface albedo and emissivity. The model is run for 10 years to reach a repeatable annual behavior.

The predicted latitude-season stability pattern obtained from this basic parameterization is compared to observations in Figure 3a (parameters are summarized on Table 1). A strong mismatch is evident: on pole-facing slopes, $CO_2$ ice is expected to be stable down to 22°S and should be observed over a period twice longer at 45°S. This discrepancy implies that a physical source of heat, localized on slopes, has not been taken into account. We have identified three major unknowns in our modeling approach that could significantly impact the results. First, the physical properties of the $CO_2$ ice deposits are not constrained by observations with a good accuracy. Changing the $CO_2$ ice parameters within the range of known properties (Table 1) cannot explain the mismatch, notably because ice properties do not significantly impact the starting date of the condensation. Secondly, steep pole facing slopes in winter are mainly illuminated by the light scattered by the airborne dust particles contrary to flat surfaces. To compute this scattered flux, the LMD model use a robust slope illumination scheme derived from Monte-Carlo simulations [*Spiga and Forget*, 2008]. However, uncertainties remain in the optical properties of dust particles, as well as in the amount of dust. In addition, the flux scattered toward the slope when the sun is below the horizon is not routinely accounted for because radiative transfer calculations are performed using a plane-parallel geometry. Some examples of predictions of the model using different hypotheses for the scattered light contribution are shown on Figure 3a. Increasing the amount of scattered light within the uncertainty range makes it possible to reduce the stability extent of $CO_2$ ice by 5° of latitude and 50° of solar longitude, which is not enough to explain the mismatch.

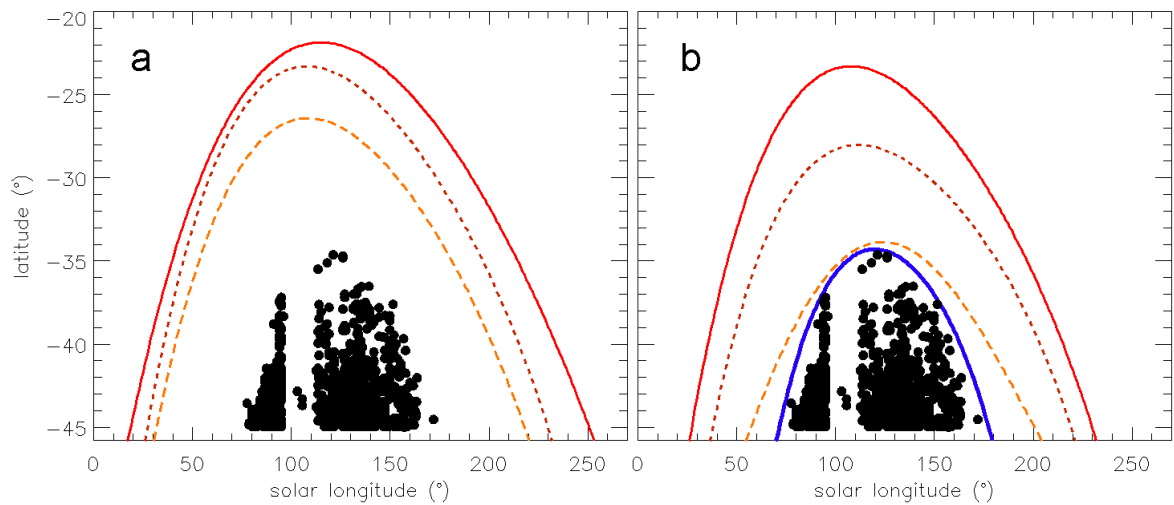

*Figure 3: Observed $CO_2$ ice deposits (black points) are shown in a latitude/season diagram, and compared to modeling predictions of the stability limit (lines). The $CO_2$ deposits observed at the edges of the distribution are located on the relatively flat plateau at the east of Hellas. (a) Impact of scattered light contribution. Three dust optical depth scenarios are shown: the standard LMD GCM scenario [Forget et al., 1999] in solid line (τ ~ 0.15 in winter), the optical depth measured by Spirit [Lemmon et al., 2004] in dotted line (τ ~ 0.23 once scaled for elevation), and the Opportunity scenario in dashed line (τ ~ 0.4). Even a high winter optical depth value for the southern hemisphere (0.4) can not explain the observations. The Spirit scenario is used in Figure b as it is a good proxy for the eastern Hellas plateau [Vincendon et al., 2009]. (b) Impact of ground thermal inertia: From top to bottom, the ground is composed of a single layer with thermal inertia I = 250 kg $K^{-1}$ $s^{-5/2}$ (solid line), I = 600 kg $K^{-1}$ $s^{-5/2}$ (dotted line), I = 1180 kg $K^{-1}$ $s^{-5/2}$ (dashed line). The best fit (thick solid line) is obtained with two layers (I = 250 kg $K^{-1}$ $s^{-5/2}$ above I = 2120 kg $K^{-1}$ $s^{-5/2}$) and with a latitude dependent depth (6 cm at 45°S, 13 cm at 35°S, and an upper limit of 90 cm at 25° to extinct $CO_2$ ice). A change in the upper layer inertia significantly modifies the depth retrievals: at 45°S, the depth is 2 cm for I = 100 kg $K^{-1}$ $s^{-5/2}$ and 16 cm for I = 400 kg $K^{-1}$ $s^{-5/2}$; at 35°S, the depths are 5 cm and 40 cm respectively. Using a slope angle of 20° (instead of 30°) without subsurface ice leads to a model stability limit similar to the dotted line in panel b ($CO_2$ ice should be stable down to 28°S on 20° slopes).*

The last unknown is the subsurface thermal flux, controlled by thermal inertia. Higher thermal inertia surfaces store more heat during summer, and release this heat in winter [*Haberle et al.*, 2008]. As shown in Figure 3b, increasing the thermal inertia of the surface and subsurface allows the model to better fit the observations. Thermal inertia higher than 1180 kg K$^{-1}$ s$^{-5/2}$ is required to bring the stability of $CO_2$ ice to a level consistent with the observations. We can also notice that the inertia must be latitude-dependent (Figure 3b, dashed line), with a higher value around 1500 kg K$^{-1}$ s$^{-5/2}$ required at 45°S compared to 1180 kg K$^{-1}$ s$^{-5/2}$ at 35°S. However, the thermal inertia of the surface is typically in the 100 – 400 kg K$^{-1}$ s$^{-5/2}$ range on these slopes according to TES (*Thermal Emission Spectrometer* onboard Mars Global Surveyor) observations [*Putzig et al.*, 2005]. A surface inertia above 1180 kg K$^{-1}$ s$^{-5/2}$ for all pole-facing slopes is therefore not possible. The thermal inertias retrieved by TES have been modeled from the daily response of the heated surfaces and are therefore representative of the first few centimeters of the surface only [*Putzig et al.*, 2005], while the first meters of the subsurface affect surface temperatures [*Bandfield and Feldman*, 2008]. We have therefore modified our model to include two layers in the subsurface of different thermal inertia. This structure is expected if a permafrost layer is present below a dry regolith cover [*Hudson et al.*, 2009; *Aharonson and Schorghofer*, 2006]. The inertia of the upper layer is set to the mean value of 250 kg K$^{-1}$ s$^{-5/2}$ as constrained from TES observations, while the lower layer is set to 2120 kg K$^{-1}$ s$^{-5/2}$, the inertia of water ice at Martian subsurface temperature. Inertias of about 2000 kg K$^{-1}$ s$^{-5/2}$ are also representative of permafrost or solid bedrock [*Fergason et al.*, 2006; *Bandfield and Feldman*, 2008]. Simulations where performed assuming various depths for the high inertia layer. A depth varying with latitude from 6 cm at 45°S to 13 cm at 35°S is required to fit the observations (Figure 3b, thick solid line). This latitude-dependent depth is the counterpart of the latitude-dependent inertia that was required in the single layer approach (Figure 3b, dashed line): a shallower depth acts as a higher mean inertia by bringing the buried high inertia layer closer to the surface. Below 35°S, we can only derived the upper limit of the depth needed to extinct $CO_2$ ice: the depth increases more quickly at those latitudes and reaches 0.9 m at 25°S. The latitude dependence of the depth retrieved using our approach agrees very well with the predicted ice table depths on pole facing slopes by an independent study [*Aharonson and Schorghofer*, 2006]. In contrast mechanisms other than a water ice-rich subsurface to create a two layer thermal inertia surface are not plausible. For example a bedrock layer would need to be uniformly buried in longitude, increasingly buried with latitude, without ever being exposed as it is not seen in thermal infrared data [*Bandfiled and Feldman*, 2008; *Putzig et al.*, 2005]. A definitive argument in favor of the water ice explanation is that the presence or absence of $CO_2$ ice deposits in our dataset is not correlated with the thermal inertia of the upper layer derived with TES observations. This result is expected for a buried water ice rich layer because the ice table depth adjusts to the regolith inertia [*Bandfiled and Feldman*, 2008].

Using different modeling hypotheses within the range of uncertainties (Table 1, Figure 3) leads to the result that water ice is present within one meter of the surface on all 20-30° pole facing slope down to about 25°S (Figure 2). The relevant thermal depths probed are 2 or 3 meters. Hence, an ice rich layer that thick is implied, which leads to an estimated reservoir of perennial subsurface water ice of about 50 – 500 kg m$^{-2}$ on steep slopes. Thermal measurements of seasonal temperature variations could help to derive more precise permafrost depths [*Bandfield and Feldman*, 2008], notably at latitudes lower than 34°S where $CO_2$ frost is not observed.

The *Mars Science Laboratory* (MSL) rover, scheduled in 2011, is designed to navigate on slopes up to 30° and will land equatorward of 30° latitude [*MSL Landing Site Selection -*

*User's Guide to Engineering Constraints*, 2007]. Considering that the southern hemisphere combine the widespread occurrence of slopes with near-surface water ice and phyllosilicates deposits that constitute one of the top priority target for future exobiological experiments on Mars, near-surface water ice at mid-latitude could be accessible to the next mission to Mars. One of the four candidate landing sites selected so far for MSL, Holden crater, is indeed located at the edge of the subsurface water ice area at 26°S.

**Table 1.** Model parameters used in Figure 2 and 3[a]

| parameters | used value | explored range and/or constraints |
|---|---|---|
| ice albedo | 0.65 | 0.25 – 0.65 [*Titus et al.*, 2001; *Schorghofer and Edgett*, 2006; *Langevin et al.*, 2007] |
| ice emissivity | 1.00 | 0.8 – 1.0 [*Titus et al.*, 2001] |
| surface albedo[b] | 0.23 – 0.13 | TES [*Putzig et al.*, 2005] |
| surface emissivity | 0.95 | 0.90 – 1.00 [*Bandfield and Feldman*, 2008] |
| surface thermal inertia[b] | 250 kg K$^{-1}$ s$^{-5/2}$ | TES [*Putzig et al.*, 2005] |
| subsurface thermal inertia | 2120 kg K$^{-1}$ s$^{-5/2}$ | water ice (180K) |
| slope angle | 30° | MOLA [*Smith et al.*, 1999; *Kreslavsky and Head*, 2000] |
| wind | 20 m s$^{-1}$ | 0 – 50 m s$^{-1}$ |
| surface pressure[b] | annual variations at 37°S, 135°E | Global Climate Model [*Forget et al.*, 1999] |
| flat surface temperature[b] | annual variations at 152°E | Global Climate Model [*Forget et al.*, 1999] |
| aerosols optical depth[b] | Spirit scaled to an elevation of 1 km ( x 0.75) | TES, Spirit and Opportunity *[Forget et al.*, 1999; *Lemmon et al.*, 2004; *Vincendon et al.*, 2009] |

[a]If not otherwise indicated.
[b]Longitude-dependent.